\documentclass[final,3p,times,twocolumn]{elsarticle}

\usepackage{amssymb}
\usepackage{color}

\journal{Biochimica et Biophysica Acta - Proteins and Proteomics}

\begin{document}

\begin{frontmatter}



\title{Relaxation dynamics of a protein solution investigated by dielectric spectroscopy}


\author{M. Wolf\corref{cor1}}
\author{R. Gulich}
\author{P. Lunkenheimer}
\author{A. Loidl}
\cortext[cor1]{Corresponding author. Tel.: +49 821-5983611; fax: +49
821-5983649; \textit{Email address}:
martin.wolf@physik.uni-augsburg.de}

\address{Experimental Physics V, Center
for Electronic Correlations and Magnetism, University of Augsburg,
86135 Augsburg, Germany}

\begin{abstract}
In the present work, we provide a dielectric study on two
differently concentrated aqueous lysozyme solutions in the frequency
range from 1~MHz to 40~GHz and for temperatures from 275 to 330~K.
We analyze the three dispersion regions, commonly found in protein
solutions, usually termed $\beta$-, $\gamma$-, and
$\delta$-relaxation. The $\beta$-relaxation, occurring in the
frequency range around 10~MHz and the $\gamma$-relaxation around
20~GHz (at room temperature) can be attributed to the rotation of
the polar protein molecules in their aqueous medium and the
reorientational motion of the free water molecules, respectively.
The nature of the $\delta$-relaxation, which often is ascribed to
the motion of bound water molecules, is not yet fully understood.
Here we provide data on the temperature dependence of the relaxation
times and relaxation strengths of all three detected processes and
on the dc conductivity arising from ionic charge transport. The
temperature dependences of the $\beta$- and $\gamma$-relaxations are
closely correlated. We found a significant temperature dependence of
the dipole moment of the protein, indicating conformational changes.
Moreover we find a breakdown of the Debye-Stokes-Einstein relation
in this protein solution, i.e., the dc conductivity is not
completely governed by the mobility of the solvent molecules.
Instead it seems that the dc conductivity is closely connected to
the hydration shell dynamics.

\end{abstract}


\end{frontmatter}

\section{Introduction}
Proteins are essential for life. They are the building blocks of
cells and they are part of virtually every biological process
\cite{Alberts2005, Bone1992}. As enzymes they catalyze chemical
reactions and in cell membranes they build ion channels and pumps;
they are responsible for signal generation and transmission and they
also act as antibodies, hormones, toxins, anti-freezer, elastic
fibers or source of luminescence. This ubiquity of proteins since
long has triggered scientists' demand for a deeper understanding of
their structure and functionality. In organisms, proteins with
biological functions usually exist in solution and many of their
physical and functional properties are strongly influenced by the
solvent \cite{Gregory1994a}. Therefore it is vital to examine
proteins within their common environment. Due to its importance for
obtaining a deeper understanding of biological processes, the
dynamics of proteins in general is a very active field of research
\cite{Cametti2011, Doster1989, Kay1989, Frauenfelder1991, Ban2011},
which is often focused on the protein-water interaction
\cite{Pethig1992, Chen2006, Frauenfelder2009,
Doster2010a, Jansson2011}. Here a suitable and commonly employed
experimental method is dielectric spectroscopy \cite{Cametti2011,
Pethig1992, Oncley1938, Oncley1942, Grant1962, Grant1966, Essex1977,
Pethig1987, Nandi1998, Bonincontro1999, Knocks2001, Feldman2003,
Oleinikova2004}.

\begin{figure}[h]
\centering
\includegraphics[width=8cm]{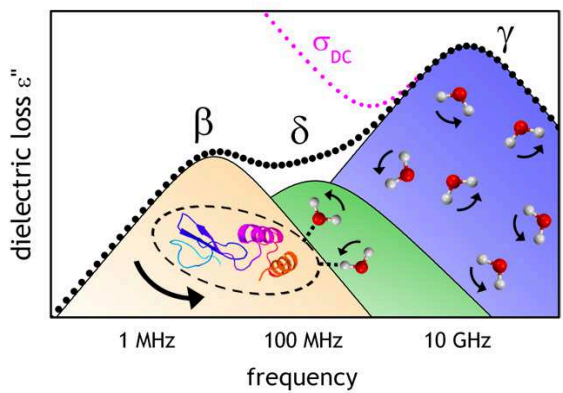}
\caption[broad]{\label{fig:schematic} Schematic view of the
dielectric loss spectrum of a typical protein solution close to room
temperature (circles). The shaded areas show the contributions of
the $\beta$, $\delta$, and $\gamma$ relaxations, which arise from
reorientational motions of the protein molecules and the bound and
free water molecules, respectively. The dotted line indicates
typical raw data, dominated by dc conductivity at low frequencies.}
\end{figure}

Dielectric spectra of aqueous protein solutions show at least three
dispersion regions \cite{Bone1992, Hasted1973, Grant1978,
Pethig1979}, revealing the typical signatures of relaxation
processes, namely a step in the dielectric constant,
$\varepsilon'(\nu)$, and a peak in the dielectric loss,
$\varepsilon''(\nu)$ (Fig. \ref{fig:schematic}). In the biophysics
community, they are often termed $\beta$-, $\gamma$-, and $\delta$-
relaxation. This nomenclature should not be confused with that used
in the investigation of glassy matter, where the terms $\alpha$-,
$\beta$-relaxation, etc. are commonly applied to completely
different phenomena than those considered here (see, e.g., refs.
\citenum{Lunkenheimer2000} and \citenum{Kremer2002}). In the present
work we follow the biophysical nomenclature.

The $\beta$-relaxation in the low frequency range and the
$\gamma$-relaxation at around 18 GHz (at room temperature) can
unambiguously be assigned to the rotation of the polar protein
molecule in its aqueous medium and the reorientational motion of the
free water molecules (similar to the main relaxation process in pure
water), respectively \cite{Grant1978, Pethig1979}. (The
$\gamma$-relaxation corresponds to the $\alpha$-relaxation of bulk
water within the glass-physics nomenclature. Within this
nomenclature, the $\beta$-relaxation may be regarded as
$\alpha$-relaxation of the protein molecules, governed by the
solvent dynamics.) The third dispersion, located between $\beta$-
and $\gamma$-relaxation is still a subject of discussion. After the
pioneering works of Oncley revealed the presence of $\beta$- and
$\gamma$-relaxations in the late 1930's and 1940's \cite{Oncley1938,
Oncley1942, Oncley1943}, first indications for this
$\delta$-dispersion where found by Haggis and Buchanan about one
decade later \cite{Haggis1951, Buchanan1952}. It is quite generally
accepted nowadays that this dispersion, which was detected in
different protein solutions, is due to bound water relaxation
\cite{Grant1962, Grant1966, Schwan1957, Schwan1965, Grant1974}. It
is well known that proteins possess a hydration shell of bound water
molecules and it is reasonable that these water molecules should
have slower dynamics than free molecules. However, the complexity of
proteins makes it difficult to decide if the $\delta$-dispersion can
solely \cite{Sun2004} be explained by a bound water relaxation
(which could also be bimodal \cite{Essex1977, Bone1992, Cametti2011,
Grant1986}) or if additional effects like intra-protein motions have
to be included \cite{Essex1977, Pennock1969, Bone1992,
Oleinikova2004, Grant1978}. The bound water relaxation has also been
discussed  in the context of the glass transition in proteins
\cite{Shinyashiki2009}.

At low frequencies ($<1$~kHz), the spectra of aqueous protein
solutions are dominated by electrode polarization (EP)
\cite{Oncley1938, Feldman2003, Emmert2011} giving rise to giant
values of the dielectric constant and a strong drop of conductivity
towards low frequencies. EP arises when the conducting ions in the
sample arrive at the metallic electrodes and accumulate in thin
layers immediately beneath the sample surface forming a so-called
space-charge region. However, this effect is not a special feature
of protein solutions, but affects dielectric spectra of any material
containing free ions \cite{Emmert2011, Wolf2011, MacDonald1987}. For
this reason, here we will not present the low frequency region of
our broadband spectra, dominated by EP contributions (see ref.
\cite{Emmert2011} for a detailed treatment of EP, including a
lysozyme solution) .

In the present work, we provide a thorough dielectric
characterization of the relaxational processes in lysozyme
solutions. Lysozyme, a representative of globular water-soluble
proteins, is an enzyme and part of the innate immune system with a
molar weight of 14.3~kDa \cite{Canfield1963, Takashima1993}. The
obtained spectra in the frequency range from 1~MHz - 40~GHz allow
for the detection of $\beta$-, $\gamma$-, and $\delta$-relaxation.
For the first time, we investigate the temperature dependence of
spectra covering all these intrinsic relaxations. This allows
gathering valuable information including, e.g., the hindering
barriers for the involved molecular motions.
Further parameters as the dipolar moment and the radius of the
protein are deduced from the dielectric results and the validity of
the Debye-Stokes-Einstein formula in this protein solution is
checked.


\section{Materials and methods}
\label{matmet} The complex dielectric permittivity and conductivity
were determined using two different coaxial reflection techniques
\cite{Boehmer1989,Schneider2001}. In the frequency range 1~MHz -
3~GHz, an Agilent Impedance/Material Analyzer E4991A, was employed.
The ac voltage is applied to a platinum parallel-plate capacitor
containing the sample material (diameter 4.8~mm, plate distance 0.1
- 0.85~mm). The capacitor is connected to the end of a coaxial line,
thereby bridging inner and outer conductor. Contributions of the
coaxial line and connectors were corrected by a calibration with
three standard impedances. For temperature-dependent measurements
the capacitor is mounted into a N$_{2}$-gas cryostat (Novocontrol
Quatro). The sample holder and coaxial line, which connects the
sample within the cryostat to the measuring device, were designed in
our laboratory \cite{Boehmer1989}. The high-frequency range (100~MHz
- 40~GHz) was covered by the Agilent "Dielectric Probe Kit" using an
open-ended coaxial line, the so-called "Performance Probe", in
combination with a Agilent E8363B Network Analyzer. Here, the line
is immersed into the sample liquid, which is kept in 50 ml plastic
tubes. The temperature was controlled by means of an Eppendorf
"Thermomixer Comfort" in combination with a 50~ml "Thermoblock" that is
 mounted on the heating plate of the Thermomixer to heat the sample tube.

Dialyzed and lyophilized Lysozyme powder from chicken egg white was
purchased from Sigma-Aldrich (Fluka 62970) and used without further
purification. Aqueous protein solutions were prepared by dissolving
weighed amounts of protein powder in deionized H$_2$O (Merck
"Ultrapur"). The investigated concentrations correspond to 3~mmol
and 5~mmol of Lysozyme powder added to one liter of water (equal to
42.9~mg and 71.5~mg per 1~ml of water). The pH values of these
solutions are around 3.8 (measured with a pH tester from
Hanna-Instruments).


\section{Results and discussion}

\begin{figure}[h]
\centering
\includegraphics[width=7cm]{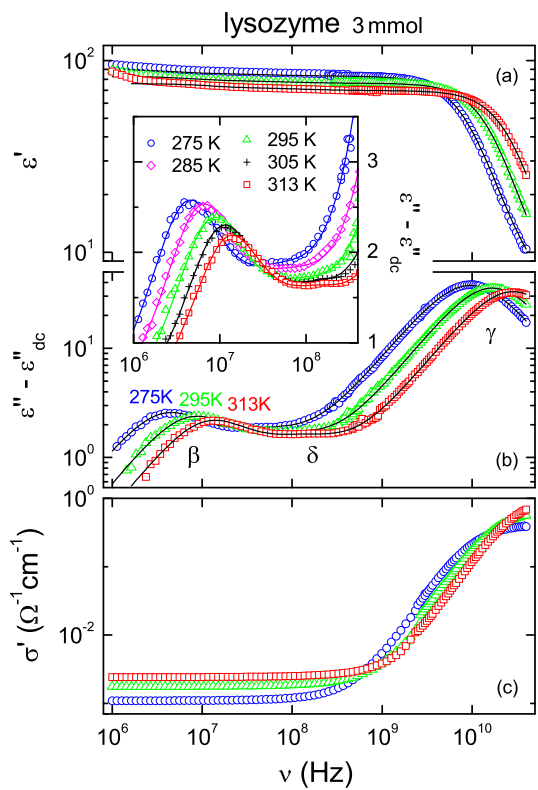}
\caption[broad]{\label{fig:broad} (a) Dielectric constant, (b)
dielectric loss (corrected for the contribution from dc
conductivity), and (c) real part of the conductivity as function of
frequency, measured at different temperatures. The lines are fits
using the sum of a Debye function for the $\beta$-relaxation and two
Cole-Cole functions for the $\delta$- and $\gamma$-relaxations.
Inset: magnified view of the dielectric loss (b) in the region of
$\beta$- and $\delta$-relaxation for five temperatures.}
\end{figure}
Figure \ref{fig:broad} shows the spectra of dielectric constant
$\varepsilon'(\nu)$ (a), dielectric loss $\varepsilon''(\nu)$ (b),
and real part of conductivity $\sigma'(\nu)$ (c) of a 3~mmol
lysozyme solution covering the frequency range from 1~MHz - 40~GHz
for three selected temperatures. The inset shows a magnified view of
the loss in the region of the $\beta$- and $\delta$-relaxations for
five temperatures in a semilogarithmic plot.

In the real part of the conductivity, a strong dc-contribution is
found showing up as frequency-independent plateau from 1~kHz (not
shown) to 100~MHz (Fig. \ref{fig:broad}(c)). This is due to ionic
charge transport with the ions arising from the residual salt
content in the protein sample (mainly chloride ions, left over from
production process). Dielectric loss (b) and conductivity (c) are
directly correlated via
$\sigma'(\nu)=\varepsilon''(\nu)\omega\varepsilon_{0}$
($\omega=2\pi\nu$ is the circular frequency and $\varepsilon_{0}$ is
the permittivity of vacuum). Thus, the dc conductivity gives rise to
a contribution
$\varepsilon''_{dc}=\sigma_{dc}/(\omega\varepsilon_{0})$, i.e. a
$1/\nu$ divergence in the loss, which obscures the detection of
possible relaxation processes at low frequencies (cf. dotted line in
Fig. \ref{fig:schematic}). Therefore it is common practice to
subtract the dc contribution \cite{Ban2011, Knocks2001,
Oleinikova2004, Sun2004, Miura1994}. This is a critical task as the
amplitude of the resulting relaxation peaks can strongly depend on
the value of the subtracted dc conductivity. In the present case,
the correctness of the subtracted values, which were determined from
the measured conductivity, is confirmed by the high quality of
simultaneous fits of the step in $\varepsilon'(\nu)$ (unaffected by
the dc conductivity) and of the peak in $\varepsilon''(\nu)$
revealed after subtraction (see below). This leads to the corrected
loss spectra shown in Fig. \ref{fig:broad}(b). In this figure, clear
signatures for the $\beta$- and $\gamma$-relaxation are found. They
show up as peaks close to 10~MHz or 20~GHz, respectively, shifting
to lower frequencies with decreasing temperature. This temperature
dependence directly mirrors the reduction of reorientational
mobility of the protein and free water molecules when temperature is
lowered. However, a closer inspection of Fig. \ref{fig:broad}(b)
provides clear indications for a third relaxation process
($\delta$-relaxation) in the frequency range around 100~MHz. There
is significant excess intensity, not explainable by a simple
superposition of $\beta$- and $\gamma$-peaks. At the highest
temperature shown (313~K), there is even the indication of a
separate weak peak (see also inset of Fig. \ref{fig:broad}).
However, clearly the $\gamma$-relaxation is the dominating process
and thus the $\beta$- and $\delta$-relaxations are hardly
discernible in the real part of the dielectric constant (Fig.
\ref{fig:broad}(a)). For low frequencies, $\varepsilon'(\nu)$
approaches a plateau whose absolute value of about 80-90 is of the
same order of magnitude as the static dielectric constant of pure
water ($\varepsilon'$ = 80.3 at 293~K \cite{Kaatze1989}). The
$\gamma$-relaxation is sufficiently strong to be detected also in
$\sigma'$ (Fig. \ref{fig:broad}(c)), despite the significant dc
contribution: It leads to a strong increase at $\nu> 200$~MHz,
followed by the approach of a plateau close to the upper boundary of
the investigated frequency range.

\begin{figure}[ht]
\centering
\includegraphics[width=7cm]{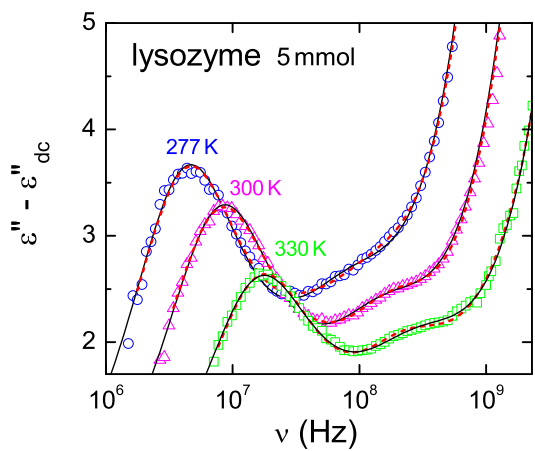}
\caption[comp]{\label{fig:comp} Dielectric-loss spectra of a 5~mmol
lysozyme solution in the region of the $\beta$- and
$\delta$-relaxations at different temperatures. The solid lines are
fits using the sum of a Debye function for the $\beta$-relaxation
and two Cole-Cole functions for the $\delta$- and
$\gamma$-relaxations. Dashed lines represent fits with four Debye
functions according to \cite{Cametti2011}.}
\end{figure}

Qualitatively similar results as shown in Fig. \ref{fig:broad} were
also obtained for a 5~mmol lysozyme solution. In Fig. \ref{fig:comp}
we show $\varepsilon''(\nu)$ within the frequency region of the
$\beta$- and $\delta$-relaxations. Obviously, in this solution with
higher protein concentration, the $\delta$-relaxation is even more
pronounced with the loss showing clear shoulders close to about
200~MHz for the two higher temperatures shown.

For a meaningful analysis of relaxational processes in dielectric
spectra, suitable fits, simultaneously performed for real and
imaginary part of the dielectric permittivity, are essential. In the
simplest case, the contributions of relaxation processes in
dielectric spectra can be fitted by the Debye equation
\cite{Debye1929}:

\begin{equation}
\label{Db}
\varepsilon^{*}(\nu)=\varepsilon_{\infty}+\frac{\Delta\varepsilon}{1+\mathrm{i}\omega\tau}\:
\end{equation}
\noindent

\noindent
$\Delta\varepsilon=\varepsilon_{\mathrm{s}}-\varepsilon_{\mathrm{\infty}}$
is the dielectric strength with $\varepsilon_{\mathrm{s}}$ and
$\varepsilon_{\infty}$ the limiting values of the real part of the
dielectric constant for frequencies well below and above the
relaxation frequency $\nu_{\mathrm{relax}}=1/(2\pi\tau)$,
respectively. At $\nu=\nu_{\mathrm{relax}}$ a peak shows up in the
dielectric loss and an inflection point in the frequency dependence
of the dielectric constant. The dc-conductivity was subtracted
before the fitting procedure. The Debye theory assumes that all
dipolar entities relax with the same relaxation time $\tau$. In
reality, however, a disorder-induced distribution of relaxation
times often leads to a considerable smearing out of the spectral
features \cite{Sillescu1999, Ediger2000}. An appropriate
phenomenological description is given by the Havriliak-Negami
formula, which is an empirical extension of the Debye formula by the
additional parameters $\alpha$ and $\beta$ \cite{Havriliak1966,
Havriliak1967}:

\begin{equation}
\label{hn}
\varepsilon^{*}(\nu)=\varepsilon_{\mathrm{\infty}}+\frac{\Delta\varepsilon}{\left[1+(\mathrm{i}\omega\tau)^{1-\alpha}\right]^{\beta}}\:
\end{equation}

\noindent Special cases of this formula are the Cole-Cole formula
\cite{Cole1941} with $0\leq\alpha<1$ and $\beta=1$ and the
Cole-Davidson formula \cite{Davidson1950, Cole1952} with $\alpha=0$
and $0<\beta\leq1$. While the Havriliak-Negami and the Cole-Davidson
functions are purely empirical, the Cole-Cole distribution of $\tau$
can be approximately derived by the microscopic model of a Gaussian
distribution of energy barriers, leading to a symmetric broadening
(compared to the Debye case) of the relaxation peak in
$\varepsilon''$ \cite{Hochli1990}.

The $\gamma$- and $\beta$-relaxations in protein solutions are
commonly found to be well describable by Eq. (\ref{Db}) or to show
at least a behavior very close to monodispersive \cite{Cametti2011,
Knocks2001, Oleinikova2004, Grant1986} and only few authors apply
the Cole-Cole function to describe the $\beta$-process
\cite{Bonincontro1999}. For the $\gamma$-relaxation, this is
reasonable as the corresponding relaxation of pure water is also of
Debye type \cite{Kaatze1989, Collie1948, Mattar1990} or only
slightly broadened \cite{Hasted1973, Grant1974a, Schwan1976}. Also
for the $\beta$-relaxation Debye behavior can be expected as any
interaction between the protein molecules is unlikely and each
molecule "sees" essentially the same environment, dominated by the
(on the timescale of the $\beta$-relaxation) quickly fluctuating
water molecules. However, for the  $\delta$-relaxation the situation
is far from being clarified, especially as the unequivocal detection
of its spectral shape is hampered by the superposition from the
adjacent $\beta$- and $\gamma$-relaxations (Figs. \ref{fig:broad}
and \ref{fig:comp}). As mentioned above, until now it is even not
clear if there is only a single $\delta$-relaxation or if several
relaxation processes contribute in this region. In ref.
\cite{Cametti2011}, where lysozyme solutions of various
concentrations were investigated at room temperature, sophisticated
arguments favoring the use of two Debye functions to describe the
$\delta$-relaxation at high concentrations were provided. In the
present work, we fit the experimental data assuming a single peak
only, which, however, is broadened according to the Cole-Cole
equation (cf. Eq. (\ref{hn}) with $\beta=1$). When adopting the
bound-water explanation of the $\delta$-relaxation, a distribution
of relaxation times (and thus of energy barriers) seems reasonable
as there should be a variation in the strength of bonding of the
water molecules of the protein surface. This may depend on the polar
residue of the macromolecules, to which the water molecule is bound,
and it may also arise from the presence of several hydration shells,
the molecules in the outer ones being more loosely bound than those
in the innermost one \cite{Cametti2011}. Bound-water relaxations
have also been previously described by the Cole-Cole function
\cite{Shinyashiki2009, Khodadadi2008a}.

The lines in Figs. \ref{fig:broad} and \ref{fig:comp} are fits with
the sum of one Debye function for the  $\beta$- and two Cole-Cole
functions for the $\delta$- and $\gamma$-relaxations. Reasonable
fits of the experimental spectra could be achieved in this way. The
width parameter $\alpha_{\delta}$ of the $\delta$-relaxation was
found to vary only weakly around 0.1 and thus was constrained to the
range 0.09-0.11 for the final fits. For the $\gamma$-relaxation, the
deviation from the Debye case was even smaller, the maximum value of
$\alpha_{\gamma}$ being $\approx0.02$. According to Grant et al.
\cite{Grant1986}, the $\delta$-relaxation of myoglobin solutions is
bimodal (due to loosely and strongly bound water), but the
dispersion due to the loosely bound water with a peak frequency of a
few GHz can be incorporated into the $\gamma$-relaxation by using a
Cole-Cole function with distribution parameters up to $\alpha=0.07$.
In our case, having even smaller values of $\alpha$, the use of an
additional relaxation process ($\delta_2$) seems not justified,
particularly as an alternative description with a Debye function
($\alpha=0$) for the $\gamma$-relaxation does not visibly worsen the
fits. Nevertheless, for comparison, Fig. \ref{fig:comp} also shows
fits with the sum of four separate Debye functions ($\beta$,
$\gamma$, $\delta_1$, $\delta_2$) as proposed in \cite{Cametti2011,
Grant1986}. An inspection of Fig. \ref{fig:comp} by eye does
not reveal any significant differences in the quality of these fits
and, based on our data base, we believe it is not possible to make a
final decision. At least, from the viewpoint of Occam's razor, using
one relaxation process less for the description of the data seems
preferable.

\begin{figure}[h]
\centering
\includegraphics[width=7cm]{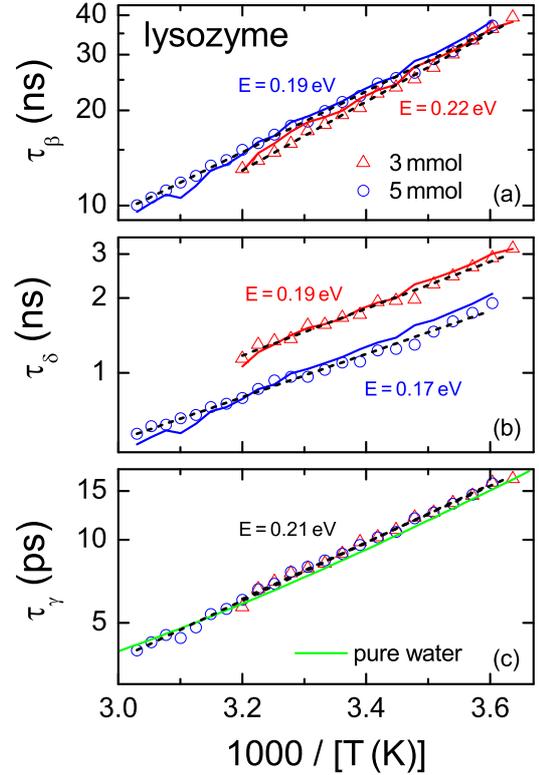}
\caption[tau]{\label{fig:tau}Temperature dependence of the
relaxation times obtained from the fitting routine. Symbols
correspond to $\tau_{\beta}$, $\tau_{\gamma}$, and $\tau_{\delta}$
of the 3~mmol and 5~mmol lysozyme solution. (a) $\beta$-relaxation;
solid lines represent $\tau_{\gamma}$, scaled
 to match $\tau_{\beta}$. (b) $\delta$-relaxation;
solid lines represent $\tau_{\gamma}$ scaled
 to match $\tau_{\delta}$. (c) $\gamma$-relaxation; the
solid line corresponds to the relaxation times of pure water after
\cite{Kaatze1989}. The dashed lines in (a) - (c) are linear fits,
from which the energy barriers were calculated according to Eq.
(\ref{tau}).}
\end{figure}


The most significant parameter obtained from an analysis of
relaxational processes is the characteristic time of the involved
dynamics of the relaxing entities. Figure \ref{fig:tau} provides the
temperature dependence of the relaxation times $\tau$ of all three
detected processes for both investigated protein concentrations. All
relaxation times reveal straight-line behavior in the Arrhenius
representation of Fig. \ref{fig:tau} (dashed lines), indicating
thermally activated behavior:

\begin{equation}
\label{tau}
\tau=\tau_{0}\exp\left(\frac{E_{\tau}}{k_{\mathrm{B}}T}\right)
\end{equation}

\noindent Here, $\tau_{0}$ is an inverse attempt frequency, often
assumed to be of the order of a typical phonon frequency and
$E_{\tau}$ denotes the hindering barriers for the relaxational
process. However, it should be noted that for all relaxation times
shown in Fig. \ref{fig:tau}, deviations from Arrhenius behavior may
well be possible when taking into account the scatter of the data
and the rather small temperature range that can be investigated in
aqueous solutions, which naturally is limited by the freezing and
boiling points of water. For pure water, such small deviations are
well documented \cite{Kaatze1989, Ellison1996}. In glass forming
liquids, an often used explanation for non-Arrhenius behavior is the
cooperativity of the molecular motions \cite{Ediger1996, Ngai2000}.
In contrast, for water also a critical power-law of $\tau(T)$
arising from a first-order phase transition was considered
\cite{Angell2008}. For comparison, in Fig. \ref{fig:tau}(c),
$\tau(T)$ of pure water from the literature \cite{Kaatze1989} is
included. The absolute values of the relaxation times of water are
close to those of the present $\gamma$-relaxation times
corroborating the assignment of this relaxation to bulk water
molecules. The two protein solutions have identical
$\gamma$-relaxation times, in accord with the findings in
\cite{Cametti2011}. Except for the highest temperatures, the
relaxation times of pure water seem to be slightly lower than those
of the protein solutions. Slower dynamics of the $\gamma$-relaxation
(i.e. higher values of $\tau$) than for pure water was also found in
other protein solutions \cite{Knocks2001, Grant1986, Mashimo1987}.

Whereas the $\gamma$-relaxation times do not depend on the protein
concentration, the $\beta$-relaxation times are slightly lower for
the solution with lower protein concentration (Fig.
\ref{fig:tau}(a)). This finding is consistent with the experimental
results of ref. \cite{Knocks2001} for ubiquitin and of
\cite{Oleinikova2004} for ribonuclease A. It evidences faster
reorientational motions of the protein molecules in the lower
concentrated sample, which may be ascribed to weaker hindering of
this motion by the neighboring molecules.

Interestingly, irrespective of the absolute values, the
$\beta$-relaxations of both lysozyme samples show nearly the same
temperature dependence as the $\gamma$-relaxation. This is
visualized in Fig. \ref{fig:tau}(a) by showing $\tau_\gamma(T)$
(solid lines), scaled to match $\tau_\beta(T)$
\cite{Frauenfelder2010, Young2011}. Thus, the $\beta$-relaxation
seems to be strongly coupled to the structural fluctuations of the
solvent (represented by the $\gamma$-relaxation). This is confirmed
by the fact that the energy barriers, determined from the Arrhenius
fits of $\tau_\beta(T)$ and $\tau_\gamma(T)$, Eq. (\ref{tau}), are
nearly identical (see Fig. \ref{fig:tau}). This coupling of $\beta$-
and $\gamma$-relaxation simply mirrors the fact that the rotation
dynamics of a molecule in a solution is essentially proportional to
the solution's viscosity, which is expressed by the Debye relation
\cite{Pethig1979}

\begin{equation}
\label{radius} \tau_{\beta}=\frac{4\pi\eta a^3}{k_{\mathrm{B}}T}\:,
\end{equation}

\noindent where $\eta$ is the solvent viscosity and $a$ is the
radius of the solute molecule. (As the factor $1/T$ in Eq.
(\ref{radius}) can be largely neglected, compared to the exponential
or even stronger-than-exponential temperature dependence usually
found for $\eta$(T), in fact Eq. (\ref{radius}) implies
$\tau_{\beta}\propto\eta$.) Just as for pure water
\cite{Angell1983}, the viscosity (translational motion) and the
$\gamma$-relaxation (reorientational motion of the free water
molecules) can be assumed to be strongly correlated in aqueous
solutions and, thus, the finding of $\tau_{\beta}\propto$
$\tau_{\gamma}$ (Fig. \ref{fig:tau}(a)) seems reasonable. Equation
(\ref{radius}) in principle allows for the determination of the
hydrodynamic radius of the protein molecules, but as they are not
exactly spherical, this can be a rough approximation only. As the
addition of such small amounts of lysozyme does not significantly
change the viscosity of water (which is proofed by the fact that the
$\gamma$-relaxation times, which are related to the sample
viscosity, are identical for the solutions and pure water), we can
use the values for pure water from \cite{Korson1969} to calculate
the hydrodynamic radius. We arrive at values of
$a=1.96~(\pm0.04)$~nm for the 5~mmol solution and $1.91~
(\pm0.02)$~nm for the 3~mmol sample, largely independent of
temperature. This is of same order as the results by Bonincontro et
al. from dielectric spectroscopy ($\approx$1.8-1.9~nm)
\cite{Bonincontro1999}, Parmar et al. and Chirico et al. from light
scattering experiments \cite{Chirico1999, Parmar2009}
(1.89$\pm$0.025~nm) and Wilkins et al. from Pulse Field Gradient NMR
\cite{Wilkins1999} (2.05~nm). In the work from Bonincontro et al.
\cite{Bonincontro1999}, a peak in $a(T)$ was found. Interestingly,
in the present data for the 5~mmol solution the faint indication of
a peak is found too, where $a$ varies between
1.93~nm and 1.99~nm. In ref.
\cite{Bonincontro1999} the peak was ascribed to the temperature
dependence of hydrophobic interactions within the protein molecules.
Also reversible denaturation effects may be considered
\cite{Sassi2011}.

As revealed by Fig. \ref{fig:tau}(b), the $\delta$-relaxation of the
5~mmol solution is clearly faster than that of the 3~mmol solution.
Interestingly, in \cite{Cametti2011} only a single Debye function
was used for the $\delta$-relaxation at low concentrations, while
for the higher ones two were necessary, the second one being located
at higher frequencies. Thus, the results in \cite{Cametti2011} may
be consistent with the present ones, namely the shift of spectral
weight to higher frequencies for higher concentrations, which in our
case is directly mirrored by the variation of $\tau$. A less
significant decrease of $\delta$-relaxation times was also observed
by Oleinikova et al. for Ribonuclease A \cite{Oleinikova2004}. To
explain this behavior, one could speculate that the structure of the
water shell around the lysozyme molecule changes in dependence of
concentration. However, our finding that the calculated radius of
the lysozyme molecule (including hydration shell) is nearly the same
for both concentrations (1.96 \textit{vs}. 1.91) speaks against such
a scenario. Thus, the microscopic origin of the observed decrease of
the $\delta$-relaxation time with increasing concentration remains
unclear.

The solid lines in Fig. \ref{fig:tau}(b) again show the
$\gamma$-relaxation times, scaled to match $\tau_{\delta}(1/T)$. At
least for the 5~mmol solution, the slope of $\tau_{\delta}(1/T)$
seems to be somewhat smaller than for $\tau_{\gamma}(1/T)$. This is
also mirrored by the energy barriers obtained from the fits with Eq.
(\ref{tau}) (dashed lines; 0.17eV for $\delta$ vs. 0.21~eV for
$\gamma$). It seems reasonable that the dynamics of the bound water
molecules is to some extent determined by the interactions with the
polar residues on the protein molecules and thus the
$\delta$-relaxation is less coupled to the structural fluctuations
of the solvent, i.e. the $\gamma$-relaxation.

The relaxation-time ratio of free and protein-bound water dynamics,
found in the present work
($\tau_{\delta}/\tau_{\gamma}=190\pm20$ for the 3~mmol
solution and $130\pm20$ for 5~mmol), is much higher than
the factor of about 6-7, reported in a recent depolarized
light-scattering study of lysozyme solutions \cite{Perticaroli2010}.
Moreover, the susceptibility spectra of that work, extending from
1~GHz well into the THz range, reveal significantly faster water
dynamics than commonly detected by dielectric spectroscopy, leading
to susceptibility peak-frequencies beyond the highest frequencies
covered in the present work. As noted in ref.
\cite{Perticaroli2010}, these light scattering experiments seem to
probe mechanisms of different physical origin than the molecular
reorientations detected by dielectric spectroscopy, which points to
even more complex water dynamics, a fact that certainly warrants
further investigation.

\begin{figure}[ht]
\centering
\includegraphics[width=7cm]{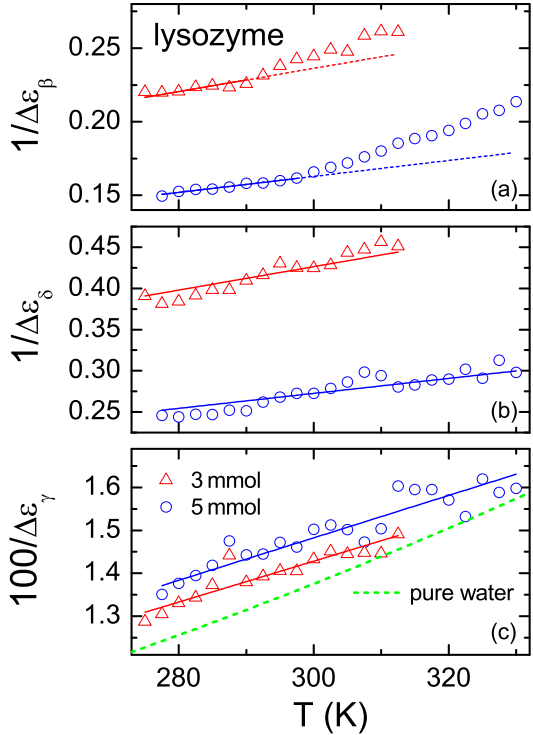}
\caption[deps]{\label{fig:deps}Temperature dependence of the inverse
relaxation strength of the (a) $\beta$-relaxation, (b)
$\delta$-relaxation and (c) $\gamma$-relaxation. Symbols correspond
to dielectric strengths evaluated by the fitting procedure. Lines
represent fits according to Eq. (\ref{curie}). The short dashed line
in (c) corresponds to $100/\Delta\varepsilon$ of pure water
according to \cite{Kaatze1989}.}
\end{figure}

Fig. \ref{fig:deps} provides information on the temperature
dependence of the relaxation strengths $\Delta\varepsilon$ of the
three observed relaxation processes. We plot the inverse
$\Delta\varepsilon$, which should result in linear behavior for
Curie behavior,

\begin{equation}
\label{curie} \Delta\varepsilon\propto\frac{1}{T}.
\end{equation}

\noindent The latter is expected for dipolar relaxations if the
Kirkwood factor, taking into account correlation effects between
dipoles, is temperature independent \cite{Boettcher1980}. Indeed the
relaxation strengths of the $\delta$- and $\gamma$-relaxations of
both concentrations can be fitted to Curie-laws (lines in Figs.
\ref{fig:deps}(b) and (c)). The relaxation strength of the
$\delta$-relaxation (Fig. \ref{fig:deps}(b)) strongly depends on the
concentration. Assuming bound water as its origin, this is a
reasonable finding because the number of bound-water molecules
should increase with the number of proteins.
$\Delta\varepsilon_{\delta}$ of the 5~mmol solution is a factor of
1.57~($\pm$0.07) higher than for 3~mmol, which is of similar
magnitude as the 5/3 ratio of the concentrations. This
proportionality is confirmed by the results of ref.
\cite{Cametti2011} (up to a concentration of 110~mg/ml, i.e.
$\approx7.7$~mmol/l), if summing up the relaxation strengths of the
two $\delta$-relaxations assumed in this work.

The absolute values of the strength of the $\gamma$-relaxation (Fig.
\ref{fig:deps}(c)) are somewhat smaller than for pure water (see
dashed line in Fig. \ref{fig:deps}(c) \cite{Kaatze1989}) and
decrease with increasing concentration. This can be primarily
ascribed to the trivial substitution effect of water by protein
molecules in the solution, i.e. the concentration of water
diminishes with increasing protein concentration. As treated in
detail in \cite{Cametti2011}, a further reduction of
$\Delta\varepsilon$ is caused by a certain amount of water molecules
being bound to the protein surface, which thus no longer contribute
to the $\gamma$-relaxation.

For the  $\beta$-relaxation (Fig. \ref{fig:deps}(a)) the
$\Delta\varepsilon_{\beta}$(T) data are consistent with Curie
behavior at the lower temperatures, at best. The deviation from
Curie behavior may be ascribed to a temperature variation of the
dipole moment of the protein molecules as discussed below.
Alternatively, a temperature-dependent Kirkwood factor may explain
the observed deviations. The $\beta$-relaxation strength of the
5~mmol solution is by a factor of 1.46($\pm0.05$) larger than for
3~mmol, which only roughly scales with the expected increase due to
the larger number density of protein molecules. A similar deviation
from a purely linear increase of $\Delta\varepsilon$ with
concentration was also found in \cite{Cametti2011} for the same
concentrations. It can be explained by a decrease of the effective
dipole moment $\mu$ of the protein molecules \cite{Cametti2011,
Oleinikova2004}.

\begin{figure}[ht]
\centering
\includegraphics[width=7cm]{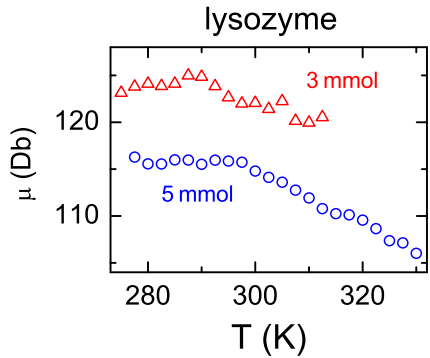}
\caption[dipmo]{\label{fig:dipmo} Temperature dependence of the
dipole moments $\mu$ as calculated from Eq. (\ref{oo}) for a 3~mmol
(triangles) and 5~mmol (circles) lysozyme solution.}
\end{figure}


In principle, $\mu$ can be calculated from $\Delta\varepsilon$.
There are a variety of time-honored models enabling such a
calculation, but as a lot of assumptions have to be made for these
models to be valid, the significance of the obtained values of $\mu$
should not be overrated. Nevertheless we used the same approach as
in ref. \cite{Cametti2011} to calculate $\mu$ by employing the
formula predicted by the Onsager-Oncley model \cite{Pethig1979,
Oncley1943}:

\begin{equation}
\label{oo}
\mu^2=\frac{2\varepsilon_{0}Mk_{\mathrm{B}}T\Delta\varepsilon}{N_{\mathrm{A}}cg_{\mathrm{K}}}
\end{equation}

\noindent Here $M$ is the protein molecular mass, $\varepsilon_{0}$
is the vacuum permittivity, $c$ is the protein concentration in
kg/m$^3$ and $k_{\mathrm{B}}$ and $N_{\mathrm{A}}$ are the Boltzmann
and Avogadro constants, respectively. $g_{\mathrm{K}}$ denotes the
Kirkwood correlation parameter, often assumed to be one in diluted
protein solutions \cite{Pethig1987}. The obtained results are shown
in Fig. \ref{fig:dipmo}. As expected from the previous paragraph, we
obtain a lower effective dipole moment for the solution with higher
protein concentration. In refs. \cite{Cametti2011} and
\cite{Oleinikova2004} this was ascribed to antiparallel correlations
between different protein molecules, implying $g_{\mathrm{K}}\neq
1$. The absolute values of $\mu$ determined from our measurements
differ from the room-temperature results reported in ref.
\cite{Cametti2011} (our values are about 1.4 times smaller). This
discrepancy remains unexplained, especially as our findings for
$\Delta\varepsilon$, used for the calculation of $\mu$, reasonably
agree with those reported in \cite{Cametti2011}. Takashima et al.
have reported a room-temperature value (extrapolated to zero
concentration) of 122~Db \cite{Takashima1993}, which is of similar
magnitude as our results. An interesting finding revealed by Fig.
\ref {fig:dipmo} is the decrease of the dipole moment with
temperature: While $\mu(T)$ is nearly constant for the lower
temperatures for both concentrations, close to room temperature it
starts to decrease with increasing $T$, which becomes especially
obvious for the 5~mmol solution, for which $\mu$ could be determined
up to higher temperatures. This directly mirrors the onset of
deviations from Curie temperature dependence of $\Delta\varepsilon$
at high temperatures, documented in Fig. \ref{fig:deps}(a). Proteins
are able to assume many nearly isoenergetic substates
\cite{Frauenfelder1991, Frauenfelder1979, Elber1987}. Thus, the
observed variation of dipole moment may well reflect gradual
conformational changes of the molecular structure at elevated
temperatures. A similar decrease of the dipole moment above about
300~K was reported for lysozyme solutions (5~mg/ml) of two different
pH values by Bonincontro et al. \cite{Bonincontro1999} and
attributed to a redistribution of microscopic state populations of
the protein. In that work, a correlation of the temperature
dependences of $\mu(T)$ and the radius $a(T)$ is assumed, while $a$
in the present study is nearly temperature independent as mentioned
above.

\begin{figure}[ht]
\centering
\includegraphics[width=7cm]{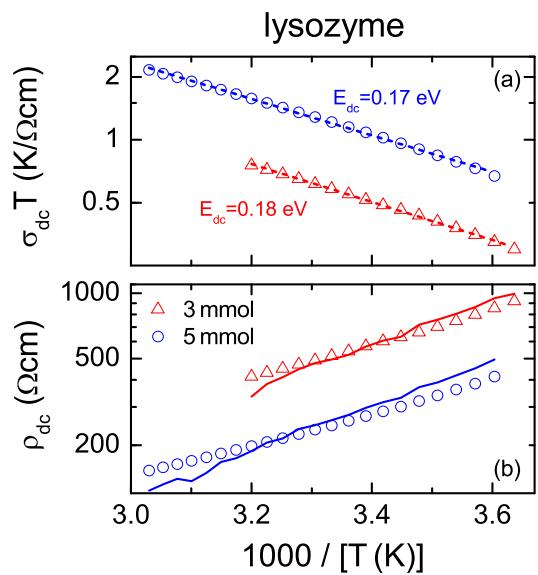}
\caption[dc]{\label{fig:dc}(a) Arrhenius type presentation of (a)
the dc-conductivity $\sigma_{\mathrm{dc}}$ of a 3~mmol (triangles)
and a 5~mmol (circles) lysozyme solution and (b) the dc-resistivity
$\rho_{\mathrm{dc}}=1/\sigma_{\mathrm{dc}}$. The dashed lines are
linear fits to the data, corresponding to Eq. (\ref{sdc}). The solid
lines represent scaled values of $\tau_{\gamma}$ to test for the
Debye-Stokes-Einstein relation,
$\rho_{\mathrm{dc}}\propto\tau_{\gamma}$.}
\end{figure}

Fig. \ref{fig:dc}(a) shows the temperature dependence of the
dc-conductivity $\sigma_{\mathrm{dc}}$ of the two solutions, plotted
in a way to linearize the Arrhenius behavior predicted for ionic
conductors, namely

\begin{equation}
\label{sdc}
\sigma_{\mathrm{dc}}=\frac{\sigma_{0}}{T}\exp\left(-\frac{E_{\sigma}}{k_{\mathrm{B}}T}\right).
\end{equation}

Here, $\sigma_{0}$ is a prefactor and $E_{\sigma}$ denotes the
hindering barrier for the diffusion of the charge carriers. Indeed
the ionic conductivity closely follows the expected thermally
activated behavior (dashed lines). The fits lead to energy barriers
of 0.18 and 0.17~eV for 3 and 5~mmol, respectively. The conductivity
is higher by about a factor of two (2.09$\pm$0.02) for the 5~mmol
solution. An increase of the conductivity is reasonable because the
ions carrying the dc current can be assumed to mainly arise from the
protein molecules releasing ions when dissolved in water. However,
it is unclear why the observed conductivity increase is higher than
the concentration ratio of 5/3.

In Fig. \ref{fig:dc}(b) the dc resistivity $\rho_{\mathrm{dc}} =
1/\sigma_{\mathrm{dc}}$ is shown and compared to the scaled
$\gamma$-relaxation times. Especially for the 5~mmol sample, clearly
different slopes of the two curves are revealed; indications for
similar  behavior is also found for 3~mmol. This finding implies a
breakdown of the Debye-Stokes-Einstein relation, which can be
expressed as $\rho_{\mathrm{dc}}\propto\tau_{\gamma}$ and is an
often discussed phenomenon in supercooled liquids \cite{Stickel1996,
Johari2006}. Interestingly this decoupling of charge transport and
$\gamma$-relaxation closely resembles the one between the $\delta$-
and $\gamma$-relaxation documented in Fig. \ref{fig:tau}(b).
Moreover, the energy barriers for dc transport (Fig.
\ref{fig:dc}(a)) and for the $\delta$-relaxation (Fig. 3(b)) are
nearly identical. This surprising finding is difficult to
rationalize and may be accidental. However, one should be aware that
the conductivity is proportional to both, the mobility and the
number of charge-carrier. If one assumes that the temperature
dependence of $\sigma_{dc}(T)$ in the present case is dominated by
the number of ions released from the protein molecules rather than
the mobility, a close connection of both processes seems possible:
For the $\delta$-relaxation, i.e. the reorientational motions of
bound water molecules, the bonds to the protein molecules have to be
(temporarily) broken, which may be determined by similar energy
barriers as those necessary for the release of ions into the
solution. Further work is necessary to corroborate this speculation,
e.g., by investigating the relation of conductivity and
$\delta$-relaxation for other protein species.

\section{Conclusions}

In the present work, we have provided a thorough characterization of
two protein solutions with different concentrations using
high-frequency dielectric spectroscopy from 1~MHz to 40~GHz. A
variety of information on the $\beta$- and $\gamma$-relaxations
arising from the protein tumbling and reorientation of the free
water molecules has been collected. Most importantly we have
detected a well-pronounced $\delta$-dispersion, attributed to bound
water dynamics, and have obtained detailed information on its
temperature dependence. Using a Debye function for the $\beta$
process, a Cole-Cole function for the $\gamma$-relaxation, and a
single Cole-Cole function for the description of the
$\delta$-dispersion, the complete broadband spectra can be well
fitted.

Temperature-dependent data on the relaxation time and strength have
been obtained for all three main dispersion regions of lysozyme
solutions enabling the determination of hindering barriers for the
relaxational processes and for the diffusion of ionic charge
carriers. Obviously all energy barriers in these protein solutions
are of similar order of magnitude, varying between 0.17 and 0.22~eV.
While we find the expected strong correlation of the $\beta$- and
$\gamma$-relaxation, the $\delta$-relaxation seems to be less
strongly influenced by the fluctuations of the solvent and, instead,
is governed by interactions with the protein molecules. We have
found a significant concentration dependence of the
$\delta$-relaxation dynamics, whose origin is unclear until now.
From our results we have deduced the hydrodynamic radius and the
temperature dependence of the dipole moment. A noticeable result is
the decrease of the latter with increasing temperatures, which also
leads to deviations of the relaxation strength of the
$\beta$-relaxation from Curie behavior. We attribute this finding to
gradual conformational changes of the protein structure.

The analysis of the temperature-dependent dc conductivity and its
comparison with the $\gamma$-relaxation time reveals a breakdown of
the Debye-Stokes-Einstein relation, i.e. the ionic charge transport
is governed by different energy barriers than the motions of the
solution molecules. Interestingly, we find that instead the charge
transport and the $\delta$-relaxation, i.e. the reorientation of
bound water molecules, are determined by identical energy barriers,
an unexpected and so far unexplained behavior.

Overall, our high-frequency dielectric measurements demonstrate the
rich dynamics of protein solutions, which shows many properties
about whose microscopic origins currently only speculations are
possible. It is clear that further work is needed, especially
covering a broad frequency range and involving temperature-dependent
measurements.





\bibliographystyle{model1-num-names}

\end{document}